\documentclass[12pt,fleqn]{article}
\renewcommand{\baselinestretch}{2.0}
\renewcommand{\baselinestretch}{1.5}

\usepackage{graphicx}

\begin{document}

\title{\Large Black Holes: Fermions at the Extremal Limit?}

\author{
   G. Ruppeiner\footnote{Electronic address: ruppeiner@ncf.edu}\\
   Division of Natural Sciences\\
   New College of Florida\\
   Sarasota, Florida 34243-2109 }

\maketitle

\begin{abstract}

I present exact results matching Kerr-Newman Black Hole thermodynamics in the extremal limit to the two-dimensional Fermi Gas.  Two dimensions are consistent with the membrane paradigm of black holes.  Key in the analysis is the thermodynamic Riemannian curvature scalar $R$, negative for most ordinary thermodynamic systems, including those near the critical point, but mostly positive for the Kerr-Newman Black Hole and the Fermi Gas.

\end{abstract}

\par
The Kerr-Newman Black Hole (KNBH), with mass $M$, angular momentum $J$, and charge $Q$, has a well-established thermodynamic structure originated by Bekenstein \cite{Bek} and Hawking \cite{Hawk}.  Logically, this should be supported by some underlying microscopic picture, a "statistical mechanics" of black holes.  But, since matter is expected to rapidly collapse to the central singularity, what entities could drive this statistics?

\par
I argue that insight results from comparing known KNBH thermodynamics to standard statistical mechanical models.  A major element here \cite{Rupp95} is the thermodynamic Riemannian curvature scalar $R$, first evaluated for the KNBH by \r{A}man, Bengtsson, and Pidokrajt \cite{Aman03}.  In the sign convention of Weinberg \cite{Weinberg}, $R$ is mostly positive for the KNBH, but mostly negative for ordinary thermodynamic systems, including most critical point models.  An exception is the three-dimensional (3D) Fermi Gas, which Janyszek and Mruga{\l}a \cite{Mrug90} found to have positive $R$.  These authors also emphasized the general importance of the sign of $R$.

\par
Here, I work out the 2D Fermi Gas and find some exact correspondences with the KNBH in the extremal limit, where the temperature goes to zero.

\par
The KNBH entropy is \cite{Dav}

\begin{equation} S(M,J,Q)=\frac{1}{8}\left(2M^2-Q^2+2\sqrt{M^4-J^2-M^2 Q^2}\right),\label{gA}\end{equation}

\noindent in geometrized units \cite{Grav}, where $S$ and $J$ are in $\mbox{cm}^2$, and $M$ and $Q$ are in cm.  The temperature $T$ is given by

\begin{equation} \frac{1}{T} \equiv \left(\frac{\partial S}{\partial M}\right)_{J,Q}=\frac{\left(K^2+2 K+L^2\right) M}{4 K},\label{gB}\end{equation}

\noindent and the heat capacity

\begin{equation} C_{J,Q} \equiv T\left(\frac{\partial S}{\partial T}\right)_{J,Q}=\frac {M^2 K(K^2+L^2+2K)}{4(L^2-2K)}. \label{gC}\end{equation}

\noindent Here, the dimensionless simplifying variables

\begin{equation} \{K,L\}\equiv\{\sqrt{1-\alpha-\beta},\sqrt{1+\alpha}\},\label{gD} \end{equation}

\noindent with $\alpha\equiv J^2/M^4$ and $\beta\equiv Q^2/M^2$.

\par
To be in the {\it physical regime} with real, positive $S$ and $T$ requires

\begin{equation} \alpha+ \beta <1. \label{gF}\end{equation}

\noindent Equality, $\alpha+ \beta=1$, has $K=T=0$ and constitutes the {\it extremal limit}, unattainable by the third law of black hole thermodynamics \cite{Carter}.  In this limit, Eqs. (\ref{gB}) and (\ref{gC}) yield

\begin{equation} C_{J,Q} =\frac {1}{16}M^3 L^2 T.\label{gF1} \end{equation}

\par Imagine now the KNBH immersed in an {\it infinite, extensive} environment.  The thermodynamic fluctuation probability is given by Einstein's formula

\begin{equation} P\propto\mbox{exp}\left(S_{tot}/k_B\right).\label{gF2}\end{equation}

\noindent Here, $S_{tot}$ is the total entropy of the universe and $k_B$ is Boltzman's constant.  If all three $(M,J,Q)$ fluctuate, $S_{tot}$ has no local maxima, and there are no stable states \cite{Lands,Rupp07}.  But, if we formally restrict one of $M$, $J$, or $Q$ as constant, reasoning it slow to fluctuate compared with the other two, stability is possible \cite{Rupp07}.

\par
Consider $(J,Q)$ fluctuating at constant $M$.  This is stable for all thermodynamic states in the physical regime.  I argued \cite{Rupp07} that with $M$ on the order of the Planck mass, such fluctuations might be physically relevant.  A straightforward exercise \cite{Rupp07,Pathria} results in the usual Gaussian approximation to Eq. (\ref{gF2}):

\begin{equation} P\propto\mbox{exp}\left\{ -\frac{1}{2}\left[ g_{22}(\Delta J)^2 + 2 g_{23}\Delta J \Delta Q + g_{33}(\Delta Q)^2\right]\right\},\label{gG}\end{equation}

\noindent with

\begin{equation} g_{\alpha\beta} \equiv -\left(\frac{8\pi}{L_p^2}\right)\frac{\partial^2 S}{\partial X^\alpha\partial X^\beta}, \label{gH}\end{equation}

\noindent $\{X^1,X^2,X^3\}\equiv\{M,J,Q\}$, $\Delta X^{\alpha}$ the deviation of $X^\alpha$ from its value at maximum $S_{tot}$, the Planck length $L_p\equiv\sqrt{\hbar G/c^3}$, and $\hbar$, $c$, and $G$ the usual physical constants.  With an infinite, extensive environment, fluctuations depend only on the black hole thermodynamics.  The environment merely sets the state about which fluctuations occur.

\par
The constant multiplier in Eq. (\ref{gH}) converts $S$ to $S/k_B$ in real units \cite{Rupp07}, essential in Eq. (\ref{gF2}).  Discussion in the black hole literature tends to focus on where $R$ is relatively large or small, and for this an overall multiplier for the metric is not particularly important.  But, a fluctuation based metric whose $R$ gets related to that in ordinary thermodynamic systems requires real units.

\par
The quadratic form in Eq. (\ref{gG}):

\begin{equation} (\Delta l)^2 \equiv g_{22}(\Delta J)^2 + 2 g_{23}\Delta J \Delta Q + g_{33}(\Delta Q)^2\label{gH11},\end{equation}

\noindent is the line element for a Riemannian geometry of thermodynamics \cite{Rupp95}.  Its physical significance is clear from Eq. (\ref{gG}): {\it the less probable a fluctuation between two states, the further apart they are}.

\par
Calculate $R$ as follows \cite{Weinberg}: the Christoffel symbols are

\begin{equation} \Gamma^{\alpha}_{\beta\gamma}=\frac{1}{2}g^{\mu\alpha}\left(g_{\mu\beta,\gamma}+g_{\mu\gamma,\beta}-g{_{\beta\gamma,\mu}} \right),\end{equation}

\noindent with $g^{\alpha\beta}$ the inverse of the metric, and the comma notation indicating differentiation.  The curvature tensor is

\begin{equation}  R^{\alpha}_{\beta\gamma\delta}=\Gamma^{\alpha}_{\beta\gamma,\delta}-\Gamma^{\alpha}_{\beta\delta,\gamma}+
\Gamma^{\mu}_{\beta\gamma}\Gamma^{\alpha}_{\mu\delta}-\Gamma^{\mu}_{\beta\delta}\Gamma^{\alpha}_{\mu\gamma},\end{equation}

\noindent and the Riemannian curvature scalar is

\begin{equation} R=g^{\mu\nu}R^{\xi}_{\mu\xi\nu}.\label{gH1}\end{equation}

\noindent $R$ is independent of the choice of coordinate system, suggesting it is a fundamental measure of thermodynamic properties.

\par
For an ordinary thermodynamic system, $|R|$ was interpreted \cite{Rupp79} as proportional to the correlation volume $\xi^d$, where $d$ is the system's spatial dimensionality and $\xi$ its correlation length; see \cite{Rupp95,John} for review.  A thermodynamic quantity, $R$, then reveals information normally thought to reside in the microscopic regime, $\xi$.  Thus, $R$ is interesting also in black hole physics, which has thermodynamic structures, but little microscopic information; see \cite{Arcioni,Aman4,Mirza} for review.

\par
By analogy, I interpret $|R|$ for black holes as the average number of correlated Planck areas on the event horizon \cite{Rupp08}.  This interpretation, putting the statistical degrees of freedom on the event horizon, fits the membrane paradigm for black holes \cite{Thorne}.

\par
For $(J,Q)$ fluctuations, Eq. (\ref{gH1}) yields

\begin{equation} R = \frac{\left(K^5+L^2 K^3-2 K^3-2 K^2+3 L^2 K-3 K+2\right)}{4\pi K \left(K^3+L^2 K-K+1\right)^2}\left(\frac{M_p}{M}\right)^2,\label{gI}\end{equation}

\noindent with Planck mass $M_p\equiv\sqrt{\hbar c/G}$.  (In geometrized units, $L_p=M_p$.)  Figure 1 shows $R$ in the physical regime.  It is real and positive, with a minimum of zero at $J=Q=0$.  $R$ is regular except at the extremal limit, where, by Eqs. (\ref{gB}) and (\ref{gI}), its limiting form is

\begin{equation}R =\frac{2 M_p^2}{\pi M^3 L^2 T}. \label{gI1} \end{equation}

\par The limiting product of curvature and heat capacity,

\begin{equation}\left(R\right)\left(\frac{8\pi}{L_p^2} C_{J,Q}\right)=\left(\frac{2 M_p^2}{\pi M^3 L^2 T}\right)\left(\frac{8\pi M^3 L^2 T}{16L_p^2}\right)=1 \label{gP},\end{equation}

\noindent is a unitless, scale free constant independent of where we are on the extremal curve.  The multiplier for $C_{J,Q}$ converts $S$ in $C_{J,Q}$ to $S/k_B$ in real units.

\par
For comparison, Table 1 reviews thermodynamic curvature in ordinary systems.  $R$ is negative where attractive interactions dominate and positive where repulsive interactions dominate.  Cases with weak interactions have $|R|$ small.  There are three somewhat quirky cases, with both positive and negative $R\mbox{'s}$; however, these are not relevant here.  (But, Cai and Cho \cite{Cai99} connected phase transitions in BTZ black holes to the "quirk" in $R$ for the Takahashi gas.)  The only known simple situation with positive $R$ diverging at low temperature is the Fermi gas, and I will turn to it for insight below.

\par
In ordinary thermodynamic systems, it is traditional to pull a constant volume $V$ out of the line element Eq. (\ref{gH11}), giving $R$ units of volume.  ("Volume" depends on dimension; for 2D systems, it is area.)  But, the KNBH has no fixed scale to pull out, and so for it we work with the full dimensionless line element Eq. (\ref{gH11}).  Its $R$ is dimensionless, and is interpreted \cite{Rupp08} as the number, rather than the volume, of correlated volumes.  For ordinary thermodynamic systems, whether or not the fixed $V$ is left in the line element makes no physical difference.

\par
For the cold 3D Fermi Gas, $R$ seems to diverge \cite{Mrug90} as $T^{-3/2}$, and not as $T^{-1}$ in Eq. (\ref{gI1}) for the KNBH.  This motivates me to work out the 2D Fermi Gas.  By the reasoning leading to Eq. (8.1.3) of \cite{Pathria}, the 2D Fermi gas has thermodynamic potential

\begin{equation} \phi(1/T,-\mu/T)=p/T=k_B g\lambda^{-2}f_2(\eta), \end{equation}

\noindent with pressure $p$, $\eta\equiv\mbox{exp}(\mu/k_B T)$, chemical potential $\mu$, thermal wavelength $\lambda\equiv h/\sqrt{2\pi m k_B T}$, particle mass $m$, weight factor $g\equiv(2s+1)$, particle spin $s$, and

\begin{equation} f_l(\eta)\equiv\frac{1}{\Gamma(l)}\int_0^\infty\frac{x^{l-1}dx}{\eta^{-1}e^x+1}. \label{gK}\end{equation}

\noindent I use obvious fluid units for all quantities, including $S$ and $T$.  The integral in Eq. (\ref{gK})  converges for $f_2(\eta)$, and yields $f_1(\eta)=\mbox{ln}(1+\eta)$.  $f_0(\eta)$ and $f_{-1}(\eta)$ follow from $f_1(\eta)$ using the recurrence relation $f_{l-1}(\eta)=\eta f'_l(\eta)$ \cite{Pathria}.

\par Define the heat capacity at constant particle number $N$ and constant area $A$ by

\begin{equation} C_{N,A} \equiv T\left(\frac{\partial S}{\partial T}\right)_{N,A}=N k_B\left[2 f_2(\eta)/f_1(\eta)-f_1(\eta)/f_0(\eta)\right].\end{equation}

\noindent The second equality is by Problem 8.10.ii of \cite{Pathria}.  The methods of \cite{Pathria} now yield the limiting low $T$ expression

\begin{equation} \frac{C_{N,A}}{Ak_B}=\frac{2\pi^3 g m k_B T}{3 h^2}. \label{gN}\end{equation}

\par Evaluating $R$ with Eq. (6.31) of \cite{Rupp95} yields

\begin{equation} R=-g^{-1}\lambda^2\left\{\frac{-2 f_2(\eta ) f_0(\eta )^2+f_1(\eta )^2 f_0(\eta )+f_{-1}(\eta ) f_1(\eta ) f_2(\eta )}{\left[f_1(\eta )^2-2 f_0(\eta ) f_2(\eta )\right]^2}\right\}.\end{equation}

\noindent Numerical evaluation over the physical range $-\infty<\mu<+\infty$ and $0<T<\infty$ indicates $R$ is always positive.  The methods of \cite{Pathria} yield the limiting low $T$ expression:

\begin{equation}R =\frac{3h^2}{2\pi^3 g m k_B T}. \label{gM}\end{equation}

\par The limiting $T$ dependences of  $C_{N,A}$ and $R$ match the corresponding KNBH quantities Eqs. (\ref{gF1}) and (\ref{gI1}).  This connection to a 2D model is consistent with the membrane paradigm of black holes \cite{Thorne}.  Furthermore, the limiting product of curvature and heat capacity,

\begin{equation}\left(\frac{R}{A}\right)\left(\frac{C_{N,A}}{k_B}\right)=\left(\frac{3h^2}{2\pi^3 g m k_B T A}\right)\left(\frac{2\pi^3 g m k_B T A}{3 h^2}\right)=1,\label{gO} \end{equation}

\noindent is a unitless, scale free constant independent of density.  The factor $A$ below $R$ undoes the traditional pulling out of $A$ in the ordinary thermodynamic line element.  $R/A$ here is analogous to $R$ for the KNBH.  The constant products Eqs. (\ref{gP}) and (\ref{gO}) are equal, remarkable for systems apparently so different.

\par
Note a key difference.  The KNBH entropy Eq. (\ref{gA}) does not go to zero in the extremal limit, as it does for the 2D Fermi Gas with its unique ground state.  Resolution probably requires a more sophisticated Fermi gas model.

\par
Results above are for $(J,Q)$ fluctuations.  Repeating the exercise for the quite different $(M,Q)$ and $(M,J)$ fluctuations \cite{Rupp08}, leads, remarkably, to the same limiting form for $R$ as Eq. (\ref{gI1}), and the same match to the 2D Fermi gas.

\newpage

\renewcommand{\baselinestretch}{1.0}
\normalsize

\begin{center}
\begin{tabular}{l|c|l}

\hline
\hline
System                                                                      &$R$ sign &Divergence          \\
\hline
1D Ising ferromagnet   \cite{Rupp81,Mrug}          & $-$      &  $T\rightarrow 0$ \\
critical region  \cite{Rupp95,Rupp79,Brody95}  & $-$       &  critical point          \\
mean-field theory \cite{Mrug}                                  & $-$      & critical point           \\
van der Waals \cite{Rupp95,Brody95}                  & $-$      &  critical point          \\
Ising on Bethe lattice \cite{Dol97}                          & $-$      & critical point           \\
Ising on 2D random graph \cite{John,Jan02}      & $-$      & critical point           \\
spherical model \cite{John,Jan03}                        & $-$      & critical point            \\
3D Bose gas \cite{Mrug90}                                     & $-$      &   $T\rightarrow 0$  \\
self-gravitating gas \cite{Rupp96}                         & $-$      &  unclear                   \\
1D Ising antiferromagnet   \cite{Rupp81,Mrug}  & $-$       &  $|R|$ small            \\
Tonks gas \cite{Rupp90B}                                      & $-$       &  $|R|$ small            \\
pure ideal gas \cite{Rupp79}                                 & $0$      &  $|R|$ small            \\
ideal paramagnet \cite{Rupp81,Mrug}                 & $0$      &  $|R|$ small            \\
multicomponent ideal gas \cite{Rupp90}             & $+$      &   $|R|$ small           \\
Takahashi gas \cite{Rupp90B}                              & +/-         &  $T\rightarrow 0$  \\
finite 1D Ising ferromagnet \cite{Brody03}           & +/-         &   $T\rightarrow 0$  \\
1D Potts model \cite{John,Dol02}                          & +/-        &  critical point           \\
3D Fermi gas \cite{Mrug90}                                    & $+$      &  $T\rightarrow 0$    \\
3D Fermi paramagnet \cite{Kav}                            & $+$      &  $T\rightarrow 0$    \\
\hline
\hline

\end{tabular}
\end{center}

\renewcommand{\baselinestretch}{2.0}
\renewcommand{\baselinestretch}{1.5}
\normalsize

Table 1.  Signs of $R$ and where it diverges for ordinary thermodynamic systems.  All signs are put into the sign convention of Weinberg \cite{Weinberg}.  A designation "$|R|$ small" means a value on the order of the volume of an intermolecular spacing or less.

FIGURE CAPTION

\par Figure 1.  $R(M/M_p)^2$ as a function of $J/M^2$ and $Q/M$ for $(J,Q)$ fluctuations.  $R$ is real, positive, and regular in the physical regime, and diverges as $T^{-1}$ at the extremal limit.

\includegraphics[width=6in]{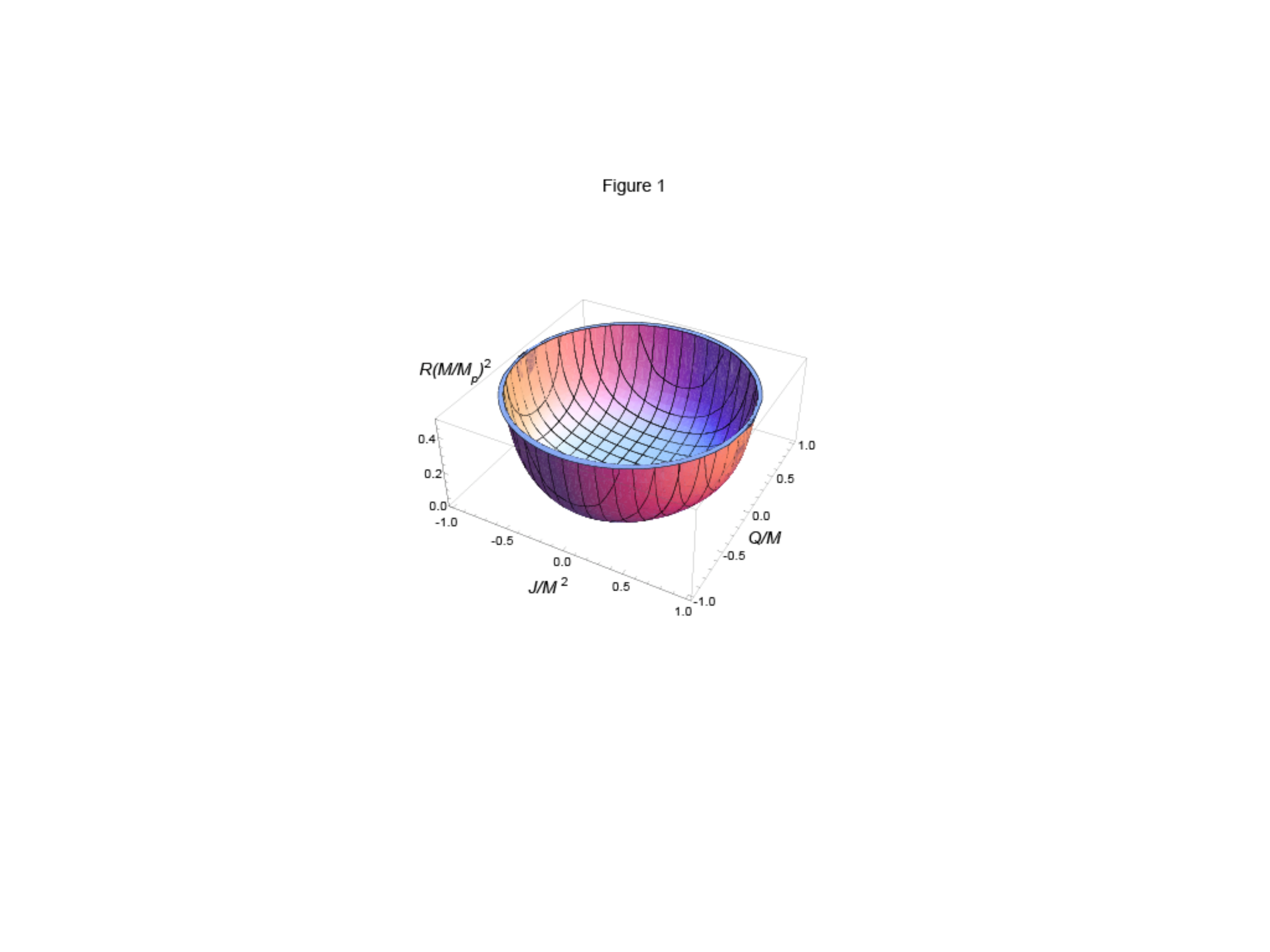}

\end{document}